\documentstyle[12pt]{article}

\begin{document}

\def\mxth{\mathsurround=0pt}
\def\xversim#1#2{\lower2.pt\vbox{\baselineskip0pt
\lineskip-.5pt \ialign{$\mxth#1\hfil##\hfil $\crcr\sim\crcr}}}
\def\gsim{\mathrel{\mathpalette\xversim >}}
\def\lsim{\mathrel{\mathpalette\xversim <}}
\def\beq{\begin{equation}}
\def\eeq{\end{equation}}
\def\noin{\noindent}
\def\grad{\bigtriangledown}
\def\hi{{\rm H}_{i}}
\def\mchi{{\rm m}_{\chi}}
\def\t{{\rm t}}
\def\tp{\acute{\rm t}}
\def\tz{{{\rm t}_{0}}}
\def\tc{{{\rm t}_{\rm c}}}
\def\lex{\langle}
\def\rex{\rangle}
\def\x{{\rm x}}
\def\chib{\bar{\chi}}
\def\k{{\rm k}}
\def\a{{\rm a}}
\def\d{{\rm d}}
\def\bx{{\bf{\rm x}}}
\def\bk{{\bf{\rm k}}}
\def\bp{{\bf {\rm p}}}
\def\i{i}
\def\phip{{\phi }_{\p} }
\def\p{\rm p}
\def\cp{\wp}
\def\pp{{\rm P}}
\def\V{{\rm V}}
\def\xba{\bar{\rm x}}
\def\h{{\rm h}}
\def\xz{{\rm x}_{0} }
\def\tr{{\rm t}_{\rm r}}
\def\lc{{{l}_{\rm c}}}
\def\M{{\rm M}}
\def\H{{\rm H}}
\def\chiz{{\chi}_{0}}
\def\pt{{\rm P}_{\rm t}}
\def\chiza{ {\chi }_{0}^{1} }
\def\chizr{ {\chi }_{0}^{\rm r} }
\def\chizs{{\chi}_{0}^{\rm s}}
\def\chia{{\chi}^1}
\def\chis{{\chi}^{\rm s}}
\def\chir{{\chi}^{\rm r}}
\def\n1{{\rm n}_{1}}
\def\ns{{\rm n}_{\rm s}}
\def\nr{{\rm n}_{\rm r}}
\def\sig {\sigma }
\def\chiz{{\chi}_{0}}
\def\chibz{{\chib}_{0}}
\def\pht{{{\hat{\rm P}}_{\rm t} }}
\def\tcri{{\rm t}_{\rm crit}}
\def\tmax{{\t}_{\rm max} }
\def\pt{{\rm P}_{\rm t}}

\titlepage

\begin{flushright} QMW-PH-93-4 \\
hep-ph/9303250\\
 \end{flushright}

\vspace{4ex}

\begin{center} \bf

DOMAIN WALL FORMATION IN THE POST-INFLATIONARY UNIVERSE\\

\rm

\vspace{4ex}

ZYGMUNT LALAK\footnote{On leave from the Institute of Theoretical
Physics, University of Warsaw. \\
$^*$BITNET: lalak@dionysos.thphys.ox.ac.uk; thomas@v2.ph.qmw.ac.uk}
\\ {\it Department of Physics \\
Theoretical Physics \\
1 Keble Road \\
Oxford OX1 3NP\\
U.K.}\\
\vspace{1ex}

and
\vspace{1ex}\\ STEVEN THOMAS \\

{\it Department of Physics \\
Queen Mary and Westfield College \\
Mile End Road\\
London E1 \\
U.K.}\\

\vspace{8ex}

ABSTRACT

\end{center}

We consider the evolution of the probability distribution $\pp
(\chi ,\chib , \t ) $, associated with an inhomogeneous light
scalar
field
$\chi $ in the Robertson-Walker Universe, where the inhomogeneities
are produced by
quantum fluctuations during an earlier  inflationary epoch. For a
specific choice of scalar potential which occurs in models of so
called late-time  phase transitions in which domain walls are produced,
 $\pp $ is shown to evolve from
a Gaussian to a non-Gaussian distribution. The structure of the
latter justifies the recent use of 3-dimensional percolation theory
to describe the initial distribution of domain walls in these
models.

\newpage
\noindent{1. $\:$ {\Large  Introduction}}
\vspace{2ex}
\\

The idea that large scale structure in the Universe might appear
 as the result of a `late-time' phase transition associated with
an extremely light scalar field, has been the subject of a number
of recent papers [1,2]. Large scale structure in these models could
 be generated either by density perturbations due to fluctuations
in this scalar field, or because of the gravitational attraction
of topological defects such as domain walls that might be produced.
If the walls form a network, then in order to avoid generating
too large a microwave background anisotropy the phase transition
should occur after photon decoupling. (Alternatively, the phase transition
could occur prior to this time and still be consistent with observations
of the microwave background if, for example, the domain walls that form
are isolated `bags' that subtend small angular distances on the sky.)
For the phase transition to occur after photon decoupling requires
the mass of the scalar to be extremely small, of order
  ${10}^{-29}$ eV, and for it to have very weak interactions [1].
Although masses of this size might at first sight appear to be
 unnatural, they have been shown to occur in several particle physics
 scenarios [2,3].

It should be understood that by `phase transition' we do not
necessarily mean one which occurs at finite temperature, which would
normally require the scalar field responsible for wall formation to
be in thermal equilibrium. Since we are dealing with a weakly interacting
field, this latter requirement might be problematic (but see ref.[1]
for a further discussion of this point.) In fact, the fluctuations in the
scalar field that eventually produce the walls need not be thermal in origin,
but might be the result of quantum fluctuations during an earlier
inflationary epoch [4]. This is the viewpoint we shall adopt in this paper.

It has been known for some time [5,6], that significant clustering
of galaxies
occurs at length scales of order 1Mpc. Moreover, detailed redshift
surveys [7], have revealed the presence of walls, voids and other
 gravitational attractors on scales up to 50Mpcs, and at redshifts
z$\,< \,$5 which corresponds to relatively recent times in the Universe's
 evolution. It is possible that such clustering might be explained
 by primordial fluctuations alone, such as those produced during
an earlier inflationary epoch. (In fact, such primordial perturbations
are at least consistent with the recent discovery of a microwave
background anisotropy [8], $\delta {\rm T}/ {\rm T} \approx 6 \times
{10}^{-6} $). However, it is difficult for such fluctuations to
simultaneously account for structure on both large and small scales.
A better approach might be a kind of hybrid model containing primordial
 density perturbations plus those due to the formation of late-time domain
walls, where the latter would describe the kind of structures
 discovered by the redshift survey [7].

In refs.[3,9], a detailed analysis of the distribution of domain
 walls in such a model was presented, where the walls are formed by quantum
fluctuations in the scalar field during an earlier inflationary epoch.
 Even though these fluctuations arise from inflation, they do not generate
 walls until much later on, when inflation ends and the Robertson-Walker
(R-W) phase begins.
(The idea that domain walls can be produced in this way was first investigated,
in a different context, in ref.[4]).
An important assumption of the analysis in refs.[3,9], was that 3-dimensional
percolation
theory [10], provides a good description of the domain wall network immediately
after their formation. The criterion for percolation theory to be strictly
applicable, is that
the scalar field $\chi $ describing the walls must be in either of two
degenerate
 minima in all causally connected regions of space, with
 probability $\cp $ and (1-$\cp$) respectively. By contrast, the probability
distribution
 $\pp (\chi, \chib ) $ of $\chi $ immediately after inflation ends is a
Gaussian
 centred about the value $\chib $. Since $\pp (\chi, \chib ) $
is the distribution of possible $\chi $ values throughout space,  it is clear
that we have to do more to meet the above criterion.

The purpose of this letter is to justify the use of percolation theory by
showing
 that the distribution $\pp (\chi ,\chib ) $ evolves during the R-W phase to
a non-Gaussian distribution whose maxima approach  two degenerate
 minima of the scalar potential. In this way the new distribution
 for $\chi $ comes much closer to satisfying the percolation criterion
discussed above.

The paper is organized as follows : in section 2 we briefly review some
concepts
 concerning scalar field quantization in de Sitter space and the appearance of
the
 distribution $\pp (\chi , \chib ) $. In section 3, we show how $\pp (\chi,
\chib) $
 evolves into a non-Gaussian distribution and illustrate this with  various
numerical plots.

\vspace{2ex}
\noindent{ 2. $\:$  {\Large Scalar field fluctuations in de Sitter
space} }
\vspace{2ex}
\\
\vspace{-1ex}

In this section we will briefly review some aspects of scalar field
quantization in de Sitter space, and explain how this leads to the
production of
an inhomogeneous quasi-classical field. The reader is referred to
refs.[11] for
details. First consider quantization of a scalar field $\phi $ in
the de Sitter
geometry d${\rm s}^2 = {\rm d}{\t}^2 - {a}^{2}{(\t)} {\rm
d}{\x}^2 $, where  $\x $ and t are  comoving coordinates.
The scale factor ${ a}(\t )= {\rm e}^{\hi \t } $, and $\hi $ is
Hubble's
constant during inflation.
\footnote{Here we assume that the field responsible for
inflation is not $\phi $, but some other field in another sector of the
theory.}
The equation of motion for $\phi $ is

\beq
\ddot{\phi } - { a}^{-2} (\t ) {\grad}^2 \phi + 3 {\hi}\dot{\phi}
  + \frac{\d {\rm V}}{{\rm d }\phi} = 0
\eeq 
\noin
Note that we choose the time variable $\t = \tilde{\t} - \tilde{{\t}_{\rm i}} $
where $\tilde{\t} $ is actual cosmological time lapsed, and $\tilde{{\t}_{\rm
i} }$
is the time at which inflation begins.
The fourier decomposition of $\phi $ is

\beq
\phi (\x , \t) = \int \frac{{\d}^{3}{\p }}{{(2 \pi )}^{3/2} }\{ \:
{a}_{\bp} \,{\rm e}^ {\i \bp\cdot \bx } \phip (\t) + {\rm h.c }\:\}
 \eeq 

\noin
where $\bp $ is the comoving 3-momentum. The modes $\phip (\t ) $
are given in terms of Hankel functions [11], whose explicit form
 we shall not need in this paper, and $a_{\bp}, ( {{a}^{\dagger}}_{\bp }  ) $
are
anihilation (creation) operators respectively.
In an infinite de Sitter space, the integration over  3-momenta $\bp
$ would continue
down to $\bp = 0 $. However, in the inflationary Universe
there is the
inflationary horizon whose comoving size  ${\hi}^{-1} $,
acts as an infrared cutoff on the spectrum.

Next we split  $\phi $ into two parts, one which includes averaging
over modes
with physical momentum   $ \k  \equiv \p \exp{(-\hi \t )} > {\hi} $
(where $\hi $ is the physical distance to the horizon at the start of
inflation),
 and the other a kind of
coarse grained field
smoothed out over horizon size scales,

\begin{eqnarray}
\phi (\x , \t) &= &\int \frac{{\d}^3 \p }{{(2 \pi )}^{3/2} } \Theta
(  \p -  \epsilon \hi {\rm e}^ { \hi \t }  ) \{ \: {a}_{\bp} \phip (\t )
{\rm e}^{\i \bp \cdot \bx } + {\rm h.c} \: \}\nonumber\\
 & +& \chi (\x , \t ) - \chib
\end{eqnarray} 

\noin
In eq.(3) $\chib $ is the mean value of the coarse grained field
$\chi $
during inflation, whose value we shall discuss later.  $\epsilon $ is a
 small parameter $0<\epsilon \ll 1 $, which ensures that the `filter'
function represented by the step function in eq.(3) rapidly vanishes
 for comoving momenta $\bp  \approx \hi {\rm e}^{ \hi \t }$ [11].
By solving the equations of motion for the modes $\phip
$, in the linearized approximation, one may deduce
that slowly varying fields $\chi $ satisfy a Langevin type
equation,

\beq
\dot{\chi }(\x , \t) = {\rm f }(\x , \t ) + ( \frac{1}{3
\hi }
{\rm e }^{-2 \hi \t } {\grad }^{2} \chi - \frac{\d {\rm V} }{\d \chi} )
 \eeq 

\noin
where  the noise ${\rm f} (\x ,\t ) $ is given by
\beq
{\rm f} (\x , \t ) = \int \frac{\d^3 \p}{{(2 \pi )}^{3/2} } \delta
(\,\p - \epsilon
{\rm e}^{\hi \t} \hi \,) \,\epsilon \,{\rm e}^{\hi \t } \,{\hi }^2
\{ \:{a}_{\p} \phip (\t ) {\rm e}^ {\i \p \cdot \x }+ {\rm h.c } \:\}
\eeq 

\noin
Since [11], one can also prove that the noise ${\rm f} (\x , \t) $
satisfies
$\lex {\rm f}(\bx , \t ) {\rm f}(\bx , \tp ) \rex  =$\\
$({\hi}^3 / 4 {\pi }^2   )
\delta (\t - \tp ) $, one may derive a Fokker-Planck (F-P) equation
from the
Langevin equation (4). Introducing the normalized probability
distribution
$\pp (\chi , \chib , \t) $ for the field
$\chi $

\beq
\lex \:{\rm F}( \chi (\x , \t ) - \chib  ) \:{\rex}_{\rm f}
\quad\equiv  \quad\int_{- \infty}^{+\infty} \d  \chi \,{\rm F}(\chi - \chib )
\,\pp (\chi , \chib ,\t)
\eeq 

\noin
where F is an arbitrary function, and $\lex \, {\rex}_{\rm f} $
signifies
averaging with respect to the noise f defined earlier, the
corresponding F-P equation takes the form

\beq
\frac{\partial \pp (\chi , \chib , \t) }{\partial \t } =
\frac{\partial}{\partial \chi}
( \frac{1}{3 \hi } \frac{\partial \V }{\partial \chi}
\pp (\chi ,\chib ,   \t ) ) + \frac{{\hi}^3}{8 {\pi}^2 }
\frac{{\partial}^{2}\pp (\chi ,\chib ,\t )}{\partial {\chi}^2 }
\eeq 

\noin
The boundary condition on $\pp (\chi , \chib , \t) $ at t= 0
$( \tilde{\t} = {\tilde{\t}}_{0} )$ , the
start of inflation, is taken to be $\pp (\chi ,\chib , \t) =
\delta(\chi -\chib )$. With this initial condition, $\pp
(\chi ,  \chib , \t ) $ for  $\t > 0$ can be interpreted as the
fraction  of the initial comoving
volume occupied by the coarse grained field $\chi $ at time t.

 We are interested in scalar field potentials
of the form

\beq
\V (\chi ) = {\mchi}^{2} {\M}^{2} [\, {\rm cos}(\frac{\chi}{\M} ) +1 \, ]
\eeq 

\noin
which occur in models  describing  light
domain walls [1]. Since in these models the mass scale M$\approx
\hi $, whilst $\mchi \ll \M $ we suspect that solution for the F.P.
eq.(7) is well approximated by the solution to the equation
with $\V = 0$. To see this, (following the derivations given in [11]
for general V), we
can obtain an approximate solution to the
F-P equation, with V given in eq. (8) as

\beq
\pp (\chi ,  \chib , \t ) = \frac{\sqrt{2 \pi}}{\M \Delta (\tau ) }
\exp{( - \frac{{(\chi -\chib )}^2 }{2 {\M}^2 {\Delta }^2 (\tau )} ) }
 \eeq 

\noin
with $\tau \equiv {\M \t }/{4 \pi } $, and where the dispersion
${\Delta }^2 (\tau) $ satisfies

\beq
\frac{\d {\Delta}^2 (\tau )}{\d \tau } = \frac{8 \pi {\mchi}^2 }
{3 \h{\M}^2 } {\rm cos}\xba (\tau ) {\Delta}^2 (\tau) +
\frac{{\h}^3}{\pi}
\eeq 

\noin
In eq.(10) $\h \equiv {\hi}/\M $, and $\xba (\tau ) $ is an intermediate
variable satisfying

\beq
     \dot{\xba} (\tau) = \frac{4 \pi {\mchi}^2 }{3 \h {\M}^2
}
{\rm sin }{\x (\tau )}
\eeq 

\noin
The point is that from the solution to eq.(11), i.e.

\beq
     \xba (\tau ) = 2 {\rm tan }^{-1} ( {\rm tan}
(\frac{\xz}{2})\: \exp{({\frac{4 \pi \tau {\mchi}^2 }{3 \h {\M}^2 }  } )\: )}
\eeq 

\noin
we see that since the ratio $\mchi /\M \ll 1\,\,$ [1],
x($\tau ) \approx \xz$,  for all relevant time scales during inflation,
where $\xz $ is a constant of
integration. This
allows one to solve eq.(10) directly for the dispersion

\beq
     {\Delta }^2 (\tau ) = \frac{3 {\h}^4 {\M}^2 }{8 {\pi}^2
{\mchi}^2 {\rm cos}(\xz ) } ( \:\exp {[\frac{8 {\pi} {\mchi}^2
}{3 \h {\M}^2 } {\rm cos }(\xz )\tau  ]}  -1\: )
\eeq 

\noin
Here $\Delta (\tau ) $ satisfies $\Delta ( \tau =0) = 0 $ corresponding
to
our previous  boundary condition $\pp (\chi , \chib , \t )=
\delta
(\chi - \chib ) $ at $ \t = 0 $. However, again we see the
very
small factors of ${\mchi}^2 /{\M}^2 $ in the exponent of eq.(13),
so we can expand and keep lowest order terms only. This gives

\beq
    {\Delta}^2 (\tau ) \approx \frac{{\h}^3 }{\pi} \tau + {\rm
O}(\frac{{\mchi}^2}{{\M}^2} )
\eeq 

 \noin
Hence, for the scalar potentials of relevance to theories
describing light domain walls, it is a very good
approximation to take as the solution to the F-P equation (7) the
usual Gaussian probability distribution

\beq
    \pp (\chi , \chib , \t ) = \sqrt{\frac{8 {\pi}^3 }{{\hi}^{3} \t}  }
\exp{({- \frac{2 {\pi}^{2}{( \chi - \chib )}^2  }{ {\hi }^{3} \t }  })}
\eeq 

\noin
centred about $\chi = \chib $.

Thus for example the  correlation of $\chi (\bx , \t )
$ with itself is

\beq
    \lex ( \chi (\bx , \t ) - \chib ) ( \chi (\bx , \t  ) - \chib )
\rex = \frac{ {\hi }^3 \t  }{4 {\pi}^2  } + {\rm
O} ({(\frac{\mchi}{\M} )}^2 )   \eeq 

Even if one had considered the correlation function in eq.(16) at two separate
(comoving) points $\bx , {\bf{\rm y}} $ the r.h.s. is typically only small
at extremely large physical separations $\mid \bx - {\bf{\rm y}} \mid \exp{\hi
\t }
\sim {\hi}^{-1} \exp
{\hi \t } $.
 Because
  the physical correlation length  of $\chi $ is thus typically of order the
inflationary horizon, $\chi $ can be interpreted as a quasi-classical
field on  very much smaller distance scales [11] .

\vspace{3ex}
\noindent{ 3. $\:${\Large Evolution of} $\pp (\chi , \chib)$
{\Large to a non- Gaussian probability distribution.}}

\vspace{2ex}

As we have indicated above and as is explained in ref.[4],
the fluctuations about the mean value of
$\chi $ give rise to an inhomogeneous quasi-classical field in the
Universe after  inflation has ended, and the Robertson-Walker (R-W)
phase begins. As explained in ref. [12], the interpretation of this field as
being
quasi-classical is clear when thinking in terms of physical momenta and length
scales,
as opposed to comoving quantities.
 The amplitude of the fluctuations in this field,
$\delta {\chi}^2 $, on physical length scales $l $ in the R-W
Universe is given by
\beq
\delta {\chi}^2 \equiv {\sigma}^2  =  \frac{{\hi}^2}{4
{\pi}^2 }\, {\rm ln}\, (\frac{l}{ \lc} )
\eeq 
\noin
where $\lc $ represents an effective ultra-violet cutoff on the physical
momentum
of the modes $\phip $ which contribute to the amplitude in eq.(16).
 The formula for the  dispersion $\sigma $ given in eq.(17)
above, is clearly related to the t- dependent fluctuations
(eq.(16)) produced during inflation. The origin of the t-dependence of
the r.h.s. of that equation ultimately comes from modes with physical
momenta  $\bk \sim $ $\hi \exp {- \hi \t }$.

The explicit form of eq.(17) may be understood as follows.
After inflation ends and the R-W phase begins, new
particle horizons appear, whose size are many orders
of magnitude smaller than that of the `blown-up' inflationary horizon.
In fact the presently observed Universe can fit many times over
inside the inflationary horizon. So we have to reconsider which
modes contribute to the fluctuations in $\chi $ when we observe
on
a physical length scale $l $ which is bounded by the presently observed
horizon. Obviously this last restriction means that we should only
keep modes whose physical momentum $\bk $
satisfies $2 \pi /\k \leq l $, so $l $
 acts as an infrared
cutoff on the spectrum of $\phip $. At the other extreme, modes
whose physical momentum is sufficiently large  so that $2\pi / \k \leq
{\H}^{-1}\, $ have decaying amplitudes (here $\H $ is the Hubble
constant during the R-W phase ) and do not contribute significantly
to the fluctuations. Therefore $\lc \approx {\rm H}^{-1} $.
By contrast, the modes whose wavelengths are
greater than the R-W horizon have constant amplitudes. All of these
 properties can be deduced by studying the time evolution of
the $\phip $ in R-W space.

To summarize the above, the emergence of R-W phase places effective
cutoffs on the spectrum of $\phip $. Taking these into account
exactly reproduces the amplitude  of fluctuations as given in eq.(17).

Before moving on to discuss the behaviour of these fluctuations at
later times in the R-W Universe, let us comment on the possible values of
the quantity $\chib $.(The following arguments were originally
 used in discussing the mean value of the Peccei-Quinn axion field [13]).
If the field $\chi $ had appeared after
inflation, then microphysical processes would tend to smooth the
field out over horizon volumes with, in general,  different values
of $\chi $ in different causally disconnected regions. (Here we
are
discussing an earlier epoch of the R-W phase when the
 cosine potential can be
neglected). In this situation
it would be natural to take $\chib $ to be the r.m.s. of all the
possible values between $-\pi $ and + $\pi $ namely ${\chi }_{\rm rms}
= \pi /\sqrt{3} $ [13].
If, on the other hand $\chi $ is present during inflation, (which
is
the case we consider in this paper), a quite different conclusion
holds. The point is that a single inflationary patch easily
encompasses our present Universe, so that the value of $\chib$ in
this patch is the one we see. But this value is actually arbitrary,
because whilst the average of $\chi $ over all inflationary
patches
would again be the r.m.s. value  $\pi /\sqrt{3} $, this says
nothing about the value of $\chi $ in any one specific patch.

This
result has important consequences for the formation of light domain
walls via the fluctuations described previously. When we try to
analyze the wall distribution in terms of percolation theory,
$\chib
$ is related to the percolation probability $\wp $[3]
(to be discussed later). The fact that
this is now arbitrary, leads to a  richer (and more realistic)
class of wall structures than had previously been  thought possible [9].

Associated to the  R-W space fluctuations in $\chi $
there is a corresponding probability density
 $\pp (\chi , \chib , l /\lc )$ given by
\beq
    \delta {\chi }^2\: =\: \lex {( \chi - \chib ) }^2 \rex \equiv \int_{-
\infty}^{+\infty} \d\chi \,{( \chi - \chib ) }^2 \,
\pp ( \chi , \chib , l/\lc )
\eeq 

\noin
which is just the Gaussian distribution of eq.(15) with de Sitter
time t replaced by ${\hi }^{-1} {\rm ln}(l/\lc )$. It is the
distribution function in eq.(18) that is important in determining
the probability of wall formation. In
particular wall creation becomes significant if the dispersion
  $\sig $ is comparable to M [3].

If one is to go beyond an estimation of the probability of wall
formation, and discuss the kind of structures the walls create
immediately after formation, statistical percolation theory seems
to offer some hope. For example in [9], it was shown that
assuming domain wall distribution can be described by that of
percolation clusters in a percolating network,  hierarchical matter
clustering occurs. If matter clustering leads to galaxy formation,
then percolation theory predicts galaxy clusters of specific
richness (total number of galaxies) and distribution which fits
fairly well with those catalogued by Abell [5].

Having said this, one really has to justify applying percolation
theory to this problem. In `two-colour' percolation, a lattice site
(for example ) is occupied or unoccupied with probability $\cp$, and
$1-\cp$ respectively. The analogue of this lattice in the cosmological
case, is the R-W Universe at some time t, being filled with
causally disconnected regions of size ${\H}^{-1}(\t ) $. Each of
these regions contains a value of $\chi $ so that over the whole
Universe, they are distributed according to the Gaussian $\pp
(\chi,
\chib, \l / \lc ) $ of eq.(18). Clearly $\pp (\chi , \chib , \l / \lc
)$, cannot
as it stands, give a distribution of $\chi $ values which are
just
$\pm \pi $ everywhere. In the remaining part of this paper, we will
show however, that this situation will change if we take into
account the effect of the cosine potential on the values of $\chi
$. We will then find that the Gaussian distribution will evolve
into one which is non-Gaussian, displaying peaks near the two
minima $\chi = \pm \pi $ of the potential. In this way we have a
situation which is much closer to the one relevant to
percolation theory.

To see how $\pp $ can evolve into non-Gaussian form it is simpler
to
think of a discrete set of distinct random field values $\chia (\t ) ,
.....\chir (\t ) $, which occupy  N causally separate regions at
time t. (N is like the lattice size in percolation models ). The
distribution of such values will be given by a discrete version of
the gaussian $\pp $, so long as the values of $\chis (\t ) $ for
s=1..r remain relatively constant in time. Since each $\chis (\t )
$ satisfies an equation of motion like that in (1) with V given by eq.(8)
 and scale  factor a(t)
appropriate to the  R-W Universe, the latter situation will be the
case when the friction due to the Hubble expansion is greater than
the acceleration on each $\chis $ due to the cosine potential. That is,
$\pp $ will maintain its Gaussian form for t $<\tz $, where $\tz$
is the time at which the $\chis $ begin to roll down their
potential. We shall refer to the values $\chia (\t ) ,.....\chir
(\t )
$ for t $<\tz $ as $\chiza , ...\chizr $. If $\n1 ,...\nr $ are
the
number of causal volumes occupied by the values  $\chiza ,
...\chizr
$ then the probability of finding the value $\chizs \, \,
{\rm for }\,\,{\rm  s} =1,..{\rm r}  $
is
$\,{\ns}/{\rm N} \,$.

Now let us consider what happens when $\t > \tz $. We have to consider
solutions of eq.(1), with the boundary conditions
$\chis (\t  =  {\t}_{0} ) = \chizs $ and
 ${\dot{\chi}}^{\rm s} (\t = \tz ) = 0 $,
s=1,..r . The solutions $\chia (\t ) ,.. \chir (\t ) $ are
trajectories with initial points $\chiza , ...\chizr $, which
in
general, will cross each other at some time t = $\tc > \tz $
(whose value we shall discuss later) even
if the initial values are all distinct. This is because the
equation of motion  for each $\chis $form a non-autonomous set of differential
equations which depend explicitly on time t. However, for $\tz  <$
t $<\tc $, the trajectories do not cross and we can approximate
them  as solutions to eq.(1) in the overdamped regime where the
term involving second derivatives is dropped. The resulting
equation is then easily seen to be autonomous, so this procedure is
consistent.

The point is that since the trajectories $\chis (\t ) $ are
unique
for $\tz <$ t $<\tc $, the probability of finding the value
$\chis
(\t) $ will be the same as for  t  $<\tz $ i.e. for the value
$\chizs $. This fact will allow us to
determine the functional form of $\pt $ on $\chi (\chiz ,\t )$,
where
we now revert back to continuum notation,  and $\pt $ denotes the
distribution of the t-dependent field $\chi (\chiz ,\t )$ for
$\tz <
$ t $<\tc $. That is

\beq
     \pt (\chi (\chiz , \t ), \chibz , \sig ) =
 \pp ( \chiz , \chibz ,\sig )
\eeq 

\noin
 Solving  eq.(19) for $\pt $, we find

\beq
\pt = \sqrt {\frac{2 \pi }{{\sig}^2 } } \exp{( {-\frac{2 {\M}^2}{
{\sig }^2 } [\, \frac{{\rm f} (\chi(\chiz ,\t ) )}{2 \M}
-  \frac{\chibz }{2 \M } \, {] }^{2 }} )}
 \eeq 

\noin
where

\beq
     {\rm f} (\chi (\chiz , \t ) ) = \chiz
\eeq 

\noin
i.e. f is the inverse function that maps the solution $\chi
(\chiz , \t )
\rightarrow \chiz $. Here we see the importance of requiring
the trajectories be unique, otherwise the inverse mapping f would
not exist for all initial values $\chiz $.

Finally, we have to take care of the  correct normalization of $\pt $.
Eq.(19) was motivated from the discrete approach, whereas
in passing to the  corresponding continuum expressions we have to
worry about Jacobian factors in going from $\pp (\chiz )
\rightarrow \pt (\chi (\t ) ) $. The latter involves changing
variables from $\chiz   = {\rm f}(\chi (\t ) ) $ to $ \chi (\t ) $

\beq
    \int_{- \infty }^{\infty } \d \chiz \, \pp ( \chiz , \chibz , \sig ) =
1 =   \int_{- \infty }^{\infty } \d \, {\chi (\t ) }\,\frac{\d {\rm f}}{\d
\chi (\t ) }  \,\pt (\chi (\t ) , \chibz , \sig )
 \eeq 

\noin
Hence, the correctly normalized continuum distribution is

\beq
    \pht (\chi (\t ) , \chibz , \sig ) =
\sqrt{\frac{2 \pi }{{\sig }^2 }}\, \frac{\d {\rm f} (\chi (\t ) ) }
{\d \chi (\t ) } \,\exp{( - \frac{1}{2 {\sig }^2 }
{[ \, {\rm f}(\chi (\t ) ) - \chibz \, ] }^2    )}
  \eeq 

\noin
To see what $\pht $ looks like, we need to solve eq.(1) in the
overdamped approximation, and then compute the inverse function f.
The solution for
$\chi (\t ) $ satisfying the boundary conditions discussed
earlier,
is

\beq \chi (\t) = 2 \M  {\rm tan}^{-1} (\, {\rm tan }
(\frac{\chiz}{2
\M} ) {\rm e}^{ \Delta (\t , \tz )}\, )
\eeq 

\beq
       \Delta (\t , \tz ) = \frac{{\mchi }^2 }{9}  (\, \frac{1}{{\H
}^2 (\t ) } -
\frac{1}{{\H}^2 (\tz ) } \,) \eeq 

\noin
 It is
then straightforward to obtain the function f, which gives the
following $\pht $

\begin{eqnarray}
  \pht ( \chi (\t ) , \chibz , \sig )& = &\sqrt{\frac{2 \pi }{{\sig}^2  } }
\frac{ {\rm e}^ {- \Delta (\t , \tz ) }  }{ (\, {\rm cos}^{2} (\frac{
\chi (\t ) }{2 \M } ) + {\rm sin }^{2} (\frac{\chi (\t )}{2 \M } \,)
{\rm e}^{- 2 \Delta (\t , \tz ) }  )} \nonumber\\
&\times &\exp {- (\,\frac{2 \pi }{{\M}^2 } [\, {\rm tan }^{-1} (
{\rm tan  } ( \frac{\chi (\t ) }{2 \M }  ) {\rm e}^{- \Delta (\t , \tz ) } -
\frac{\chibz}{2 \M } \, {] }^{2} \, ) }
\end{eqnarray} 

\noin
Let us study the distribution $\pht $ for the simple case when
$\chibz = 0 $ i.e. when the original Gaussian is symmetric about
$\chiz = 0 $, as this will allow us to see the departure from
Gaussian behaviour. We look first for the stationary points of $\pht
$, which satisfy
\beq
    {\rm tan } (\frac{ \chi (\t )}{2\M} ) = {\rm tan }[\,
\frac{( {\rm e}^ {\Delta } - {\rm e}^{-\Delta } ) }
{4 {\M}^2 } \, {\sig }^{2 }{\rm
sin}(\frac{\chi (\t ) }{\M} ) \,]\, {\rm e}^{\Delta }
\eeq 

\noin
Obvious solutions to eq.(27) are $\chi = 0 $ mod $2 \pi $, but
there are other solutions that depend on $\Delta $ but which are
difficult to find analytically. However, let us concentrate on the
central stationary point at $\chi = 0 $. One can show that

\beq
     {\rm sign} ( \,\frac{{\d}^2 {\pht }}{\d {\chi }^2 } \,{)}_{\chi = 0}
= {\rm sign} (\, \frac{{\rm e}^{\Delta} - {\rm e}^{-\Delta} }{2 {\M}^2
} - \frac{1}{{\sig}^2 }\, )
\eeq 

\noin
Thus, at t = $\tz $, $\Delta = 0 $ and the point $\chiz = 0 $ is
clearly a maximum, as it should be since  $\pp $ is a Gaussian
centred at zero. As $\Delta $ increases with time, this situation continues
until a critical value $\tcri $ is reached , which is defined by

\beq
    {\rm e}^{{\Delta }_{\rm crit} } - {\rm e}^{-{\Delta}_{\rm crit} } =
\frac{2 {\M}^2 }{{\sig }^2 }
\eeq 

\noin
When $\Delta > $  ${\Delta }_{\rm crit } $, $\chi = 0 $ changes
from a maximum to a minimum. Now, on the other hand, $\pht $ is a
bounded function as $\chi (\t ) \rightarrow \pm \infty $ hence if
there is a minimum at $\chi  =0 $ there must exist other maxima
(at least 2 ) at some non-zero values of $\chi $. We expect that
similar conclusion holds for distributions with $\chib \neq 0 $,
which although difficult to prove analytically, is verified by
plots of $\pht $   (see figs.3 ).

Before discussing these numerical plots, we should return to the
issue concerning the existence of the crossing time $\tc $ beyond
which trajectories  $\chi (\chiz , \t ) $ with  different
initial
values $\chiz $, cross. In fig.1, numerical solutions to the full
equations of motion are plotted as solid curves for various initial
values  $\chiz $. It is clear from these, that $\tc $ exists,
and
is approximately given as $\mchi \tc \approx 3$.
Note we have not considered
initial conditions beyond $\chiz / \M = \pm \pi $ because $\pht $ is
exponentially suppressed in this regime.
By contrast, the dashed curves in  fig.2
are the analytic solutions given in eqs.(24,25). We see that for
$\mchi \t \leq  {\rm O}(1) $, the latter are a good approximation
to the numerical solutions of the complete equations of motion.
Thus, the expression for  $\pht $ of eq.(26) is only  correct for
times up to  maximum  $\tmax $, with $\mchi \tmax \approx {\rm O}(1) $.

In figs.3a-3d we have plotted the distribution ${\pp}_{0} \equiv
\pht /\sqrt{2 \pi\, /{\sig }^2 } $ for  values of ($\chib ,\sig \, / \pi \M $)
corresponding to (0,0.5), (0,0.8), ($\pi \, / 4$ , 0.5) and
($\pi \, /4$, 0.8) respectively.
 The plots are for time t
$\approx \sqrt{2}
{\mchi }^{-1}  $.
The values of $\sig /\pi\M $  are chosen so as to display the
double
maxima in $\pht $, described earlier. Such values
are
 very reasonable, because as was shown in ref. [3], the  probability
of
domain walls forming in these models can be estimated as the ratio $\sig /\pi
\M $.
 The dashed curves in figs.3 are those of the original
Gaussian distribution at t = $\tz $. It is clear from these plots
that
the effect of the cosine potential is to focus probability into the
minima of the potential  at $\chi = \pm \pi $, and we can see
roughly
 that this probability is equal to the area under the dashed curve
to
the right or left of the point $\chi = \chibz $.
These two areas play the role of the probability
$\wp $ and $(1-\wp )$ in percolation theory.

In conclusion, we have shown how the probability distribution $\pp (\chi ,
\chib ) $ generated during inflation evolves from Gaussian to non-Gaussian
behaviour during the R-W phase. This was a consequence of the cosine potential
of
$\chi $ focussing probability into the two minima at $\chi = \pm \pi $.
The amount of focussing in each minima  depends on the value of $\chib $.
These results justify the percolative approach to domain wall formation
in this model, with the important consequence that the percolation probability
$\cp $ can take any value $0 < \cp <1 $. \\

\vspace{2ex}
\noindent{\large Acknowledgements }\\
We would like to thank Burt Ovrut for many useful discussions, and
past collaborations in this subject. \\

\vspace{2ex}
\noindent{$\:$\large References }
\vspace{2ex}
\begin{description}
\item{[1]} C. Hill, D. Schramm and J. Fry, {\it Comm. on Nucl. and Part. Phys.}
{\bf 19} (1989) 25; C. Hill, D. Schramm and D. Widrow, Fermilab preprint
-PUB-89/166-T (1989).
\item{[2]} C. Hill and G.Ross, {\it Nucl.Phys. } {\bf B311} (1988) 253,
 {\it Phys. Lett. } {\bf 203B} (1988) 125; A. Gupta, C. Hill, R. Holman and
E. Kolb, {\it Phys. Rev. }{\bf D45} (1992) 441; J. Frieman, C. Hill and R.
Watkins,
Fermilab-PUB-91-324-A(1991).
\item{[3]} B.A. Ovrut and S. Thomas, {\it Phys. Lett. }{\bf 277B} (1992), 53.
\item{[4]} A. Linde and D.H. Lyth, {\it Phys. Lett.}{\bf  246B} (1990) 353;
D. H. Lyth, {\it Phys.Lett. }{\bf } (1991) .
\item{[5]} G. Abell {\it Astrophys. J. Suppl.}{\bf 3}(1958) 211.
\item{[6]} N. A. Bahcall, {\it Astrophys. J. }{\bf 232 }(1979) 689.
\item{[7]} V. de Lapparent, M. J. Geller and J. Huchra, {\it  Astrophys. J. }
{\bf 302},L1 (1986) 1986.
\item{[8]} G. F. Smoot et al, {\it Astrophys. J.}(1992) TBD.
\item{[9]} Z. Lalak, B. A. Ovrut and S. Thomas, University of Pennsylvania
 preprint UPR-0507T; Z. Lalak and B.A. Ovrut UPR-0504T (1992).
\item{[10]} For example see the review by D. Stauffer, {\it Phys. Rep. }{\bf
54} (1979) 1.
\item{[11]} A.A. Starobinsky, {\it Phys. Lett.}{\bf 117B} (1982) 175,
also in `Current Topics in Field Theory, Quantum Gravity, and Strings,'
Lecture Notes in Physics, ed H.J. DeVega and N. Sanchez, vol 246 (Springer
Heidelberg, 1986)
 p. 107; A. Goncherov and A. Linde,{\it Sov. Phys. JETP}
{\bf 65} (1987) 635; S.-J. Rey, {\it Nucl. Phys. }{\bf B284} (1987) 706.
\item{[12]} `Particle Physics and Inflationary Cosmology' by A. Linde,
Contemporary Concepts in Physics vol 5, (Harwood Academic Pubs. 1990) page 161.
\item{[13]} See for example `The Early Universe' by E. Kolb and M.S. Turner,
(Addison-Wesley Publishing, 1990) page 431 for a discussion of this point.

\end{description}
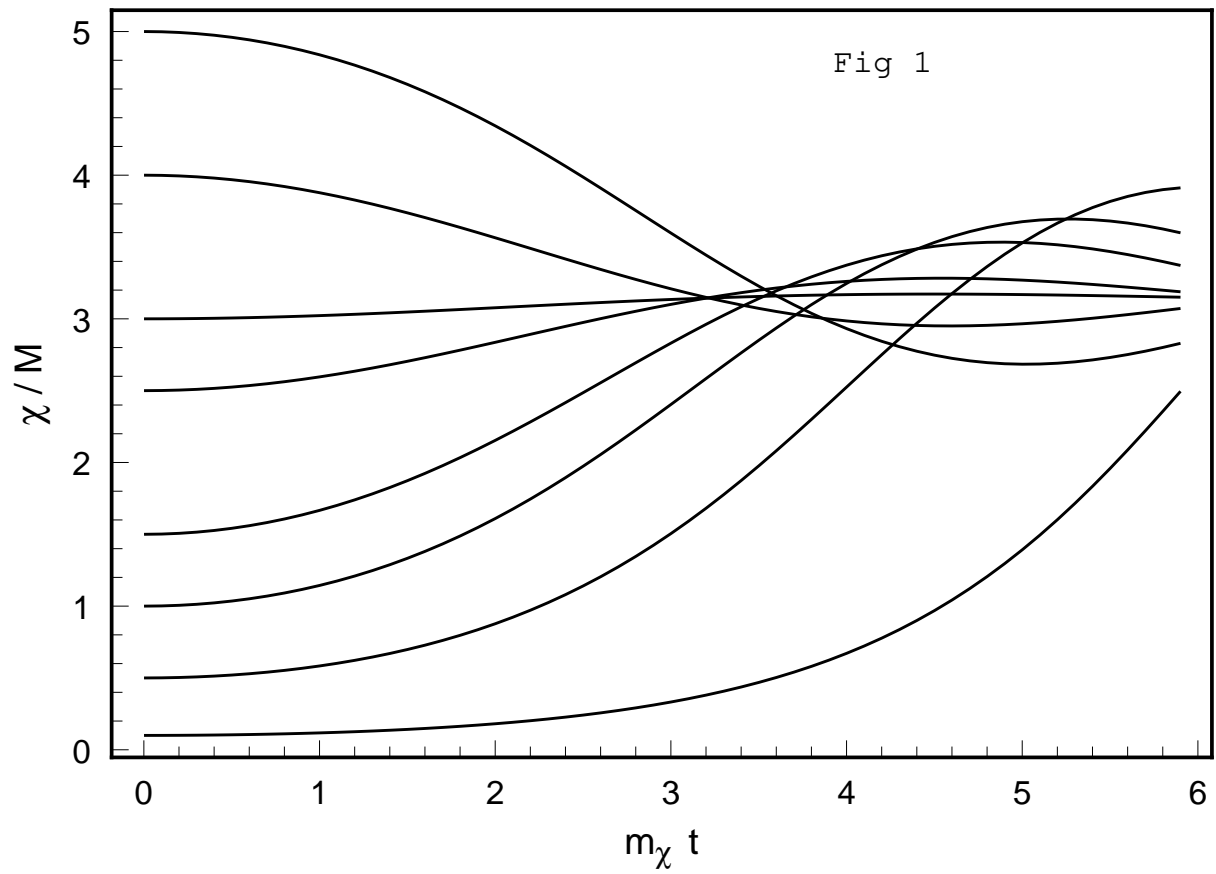
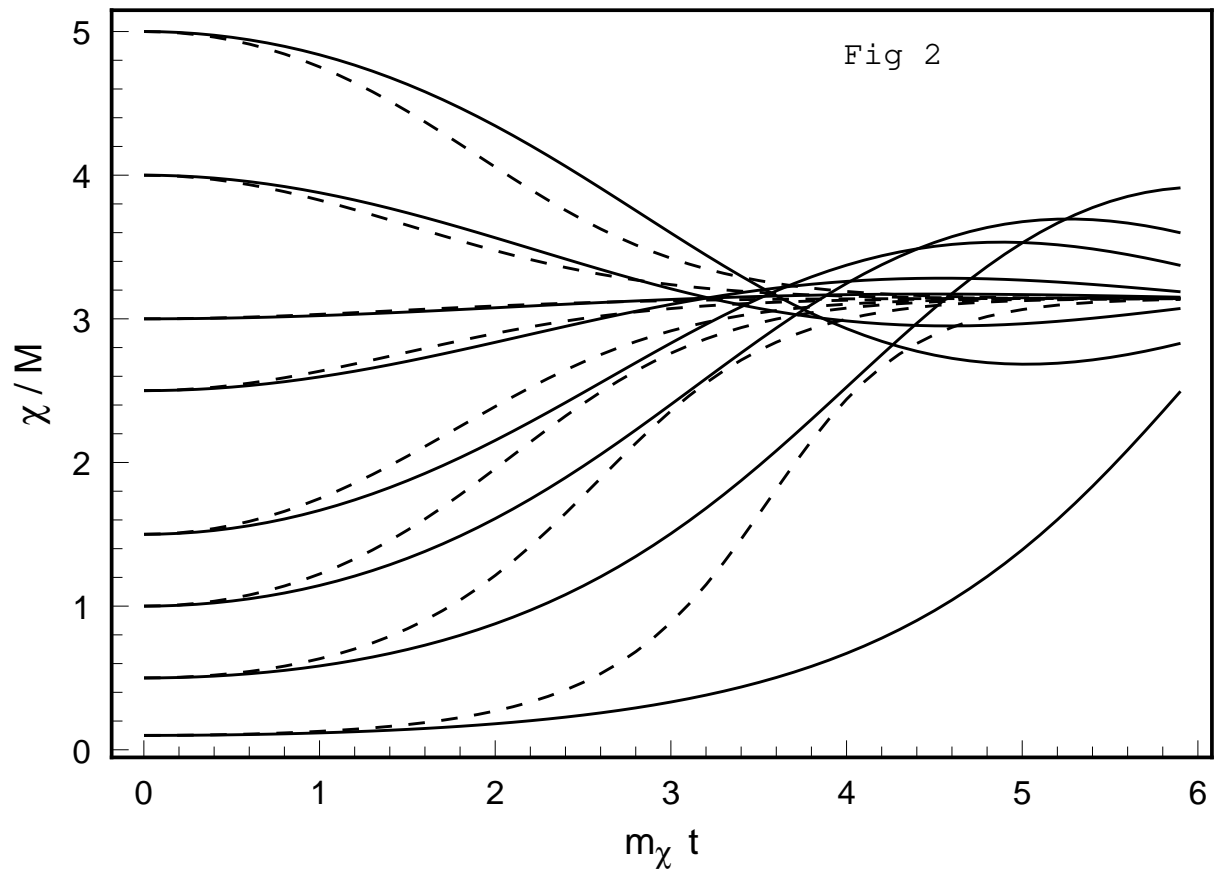
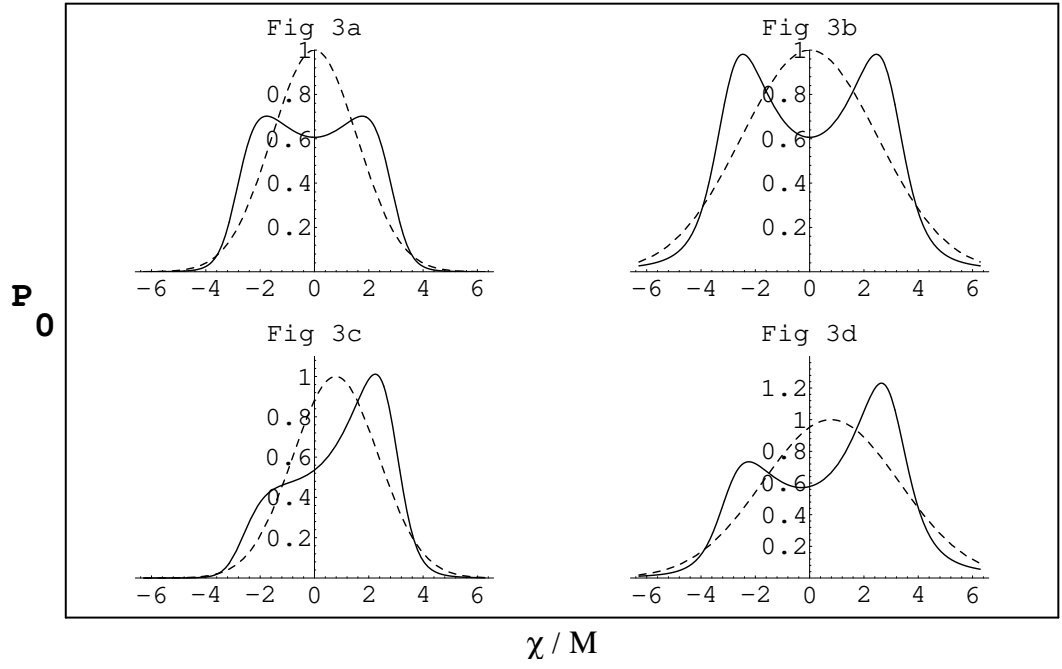
\end{document}